\documentclass[twocolumn]{aastex631}
\shorttitle{Helium in exospheres}
\shortauthors{Krishnamurthy \& Cowan}
\graphicspath{{./}{figures/}}
\received{\today}
\begin{document}

\title{Helium in Exoplanet Exospheres: Orbital and Stellar Influences}

\author[0000-0003-2310-9415]{Vigneshwaran Krishnamurthy}
\affiliation{Trottier Space Institute, McGill University, 3550 rue University,  Montr\'eal, QC, H3A 2A7, Canada}
\affiliation{Department of Physics, McGill University, 3600 rue University, Montr\'eal, QC, H3A 2T8, Canada}

\author[0000-0001-6129-5699]{Nicolas B. Cowan}
\affiliation{Trottier Space Institute, McGill University, 3550 rue University,  Montr\'eal, QC, H3A 2A7, Canada}
\affiliation{Department of Physics, McGill University, 3600 rue University, Montr\'eal, QC, H3A 2T8, Canada}
\affiliation{Department of Earth \& Planetary Sciences, McGill University, 3450 rue University, Montr\'eal, QC, H3A 0E8, Canada}



\begin{abstract}

Searches for helium in the exospheres of exoplanets via the metastable near-infrared triplet have yielded 17 detections and 40 non-detections. We performed a comprehensive re-analysis of published studies to investigate the influence of stellar XUV flux and orbital parameters on the detectability of helium in exoplanetary atmospheres. We identified a distinct `orbital sweet spot' for helium detection, 0.03 to 0.08\,AU from the host star, where the majority of detections occurred. This sweet spot is influenced by the stellar XUV flux and planet size. Notably, a lower ratio of XUV flux to mid-UV flux is preferred for planets compared to non-detections. We also found that helium detections occur for planets around stars with effective temperatures of 4400--6500\,K, with a sharp gap at 5400 to 6000\.K, where no detections occur.  Additionally, our analysis of the cumulative XUV flux versus escape velocity shows planets with helium detections are found above the `cosmic shoreline', suggesting the shoreline needs revision. The trends we found in our analysis contribute to a deeper understanding of exosphere evolution.
\end{abstract}


\section{Introduction}

In the era of exoplanetary exploration, understanding the mysteries of their atmospheres is a key goal, advancing our knowledge of these distant worlds. Since the first detection of atmosphere in an exoplanet \citep{charbonneau2002}, astronomers have utilized numerous space-based and ground-based telescopes to characterize atmospheres. 

Planets in close orbits around their star receive intense irradiation. During the initial few hundred million years following their formation, these close-in exoplanets experience atmospheric evaporation. Hydrodynamic escape plays a significant role in shaping these planets, resulting in population-level structures \citep{2003ApJ...598L.121L, 2009ApJ...693...23M, 2012MNRAS.425.2931O, 2018A&A...619A.151K}. Notable structures attributable to hydrodynamic escape include the `Neptunian desert' \citep{2013ApJ...763...12B, 2016NatCo...711201L, 2016A&A...589A..75M, 2022ApJ...924....9H, 2023ApJ...945L..36T} and `radius valley' \citep[sometimes referred to as radius gap;][]{2017AJ....154..109F, 2018AJ....156..264F, 2018MNRAS.479.4786V, 2022ApJ...941..186L}. Evolutionary models are consistent with these population-level structures \citep{2017ApJ...847...29O, 2019MNRAS.487...24G, 2020MNRAS.493..792G, 2021MNRAS.503.1526R} with few exceptions \citep[e.g.,][]{2022AJ....164..172D}, due to the timescale of loss and the age of the systems \citep[see review in][]{https://doi.org/10.1029/2020JE006639}. 

\begin{figure*}[]
  \centering
  \includegraphics[width=\linewidth]{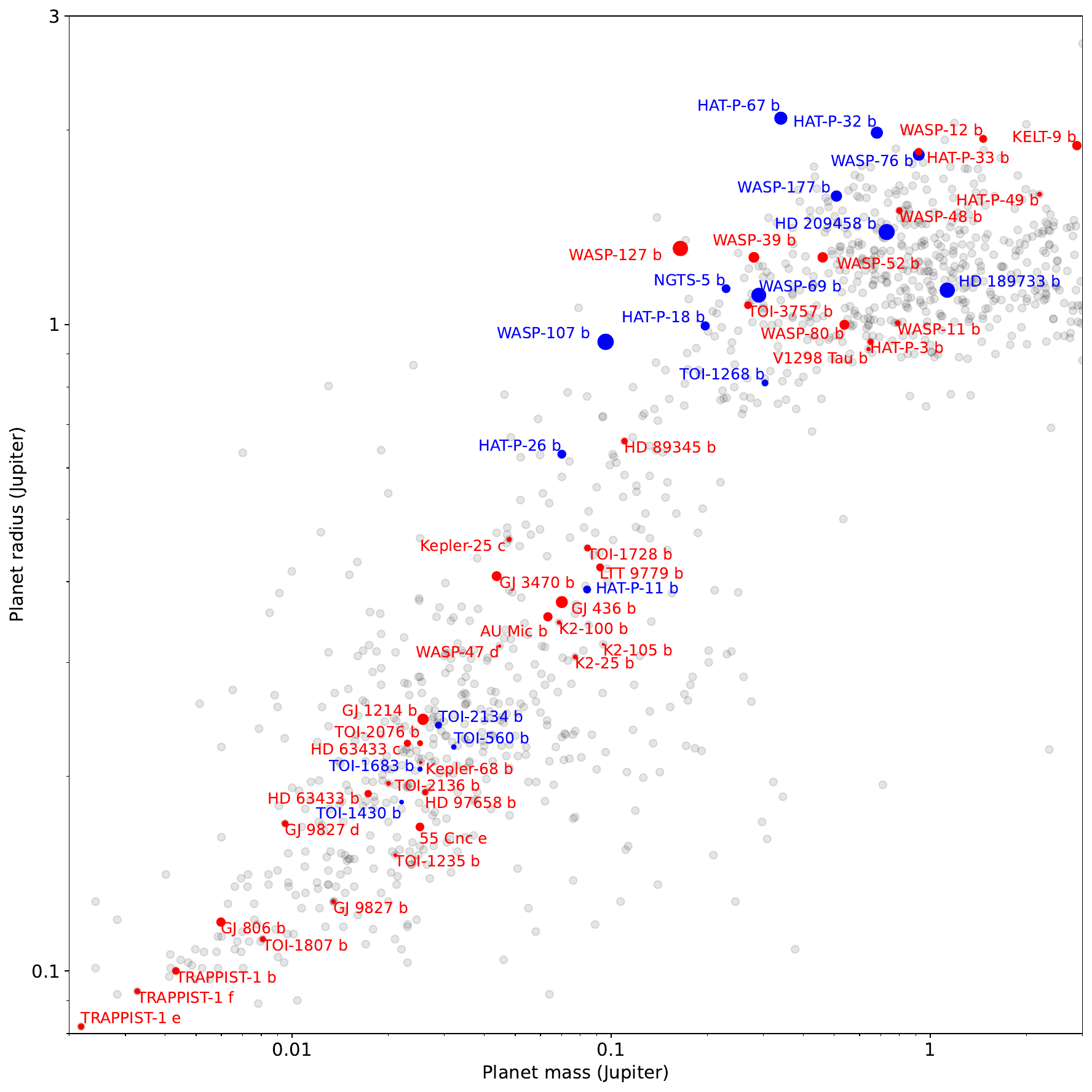}
  \caption{Radius--Mass plot for all planets from the NASA Exoplanet Archive superimposed with He\,I observed planets. Red points denote planets probed for the He\,I triplet but not detected, while blue points represent planets where the He\,I triplet was detected. The size of points has been scaled with their Transmission Spectroscopic Metric \citep[TSM;][]{tsm_eliza}. Error bars are removed for clarity.}
  \label{fig: mass-radius}
\end{figure*}

The time-dependence of atmospheric loss depends on the energy source. For photoevaporation, the star's X-ray and extreme UV (XUV) heat the planet's upper atmosphere, driving a hydrodynamic outflow, similar to a Parker wind \citep{1958ApJ...128..664P}, resulting in mass loss. XUV photons can only penetrate the upper layers of the atmosphere and are short-lived \citep{2013ApJ...775..105O}. This is mainly because the XUV flux drops with star age, although M-dwarfs maintain high XUV much longer \citep{2023AJ....166...16P}. On the other hand, in the `core-powered mass-loss' process, the upper atmosphere is heated by the infrared (IR) radiation from the cooling core of the planet \citep{2018MNRAS.476..759G}. This drives a similar hydrodynamic flow, albeit weaker and longer-lived. The population-level statistics may also be explained without invoking hydrodynamic outflows \citep{2021ApJ...908...32L, 2022ApJ...941..186L}, with delayed-degassing of hydrogen-helium from the mantle \citep{2018ApJ...854...21C} or by the `water-worlds' hypothesis \citep{2009A&A...501.1139M, 2020A&A...638A..41T, 2022Sci...377.1211L}. Although different models can explain the observed planet distributions in the mass-radii plane, they have varied implications on the planet's formation and evolution history \citep{2023MNRAS.518.4357O}.

\begin{figure*}[]
  \centering
  \includegraphics[width=\textwidth]{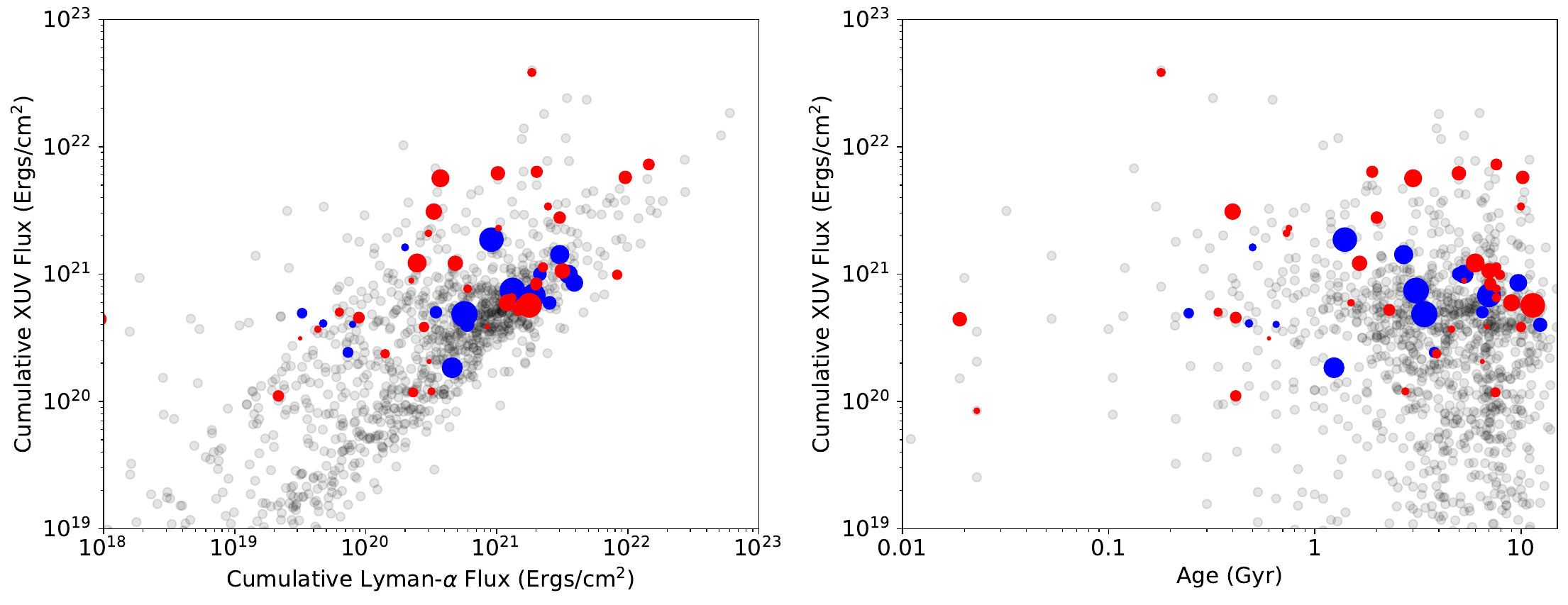}
  \caption{Cumulative XUV (X-ray + EUV) fluxes for all planets from the NASA Exoplanet Archive against the cumulative Ly-$\alpha$ flux (left) and against age (right). The blue and red TSM-scaled points denote the He\,I detected and undetected planets, respectively. It is noticeable that the helium detected planets are concentrated within a narrow range of cumulative XUV flux.}
  \label{flux_plot}
\end{figure*}

Direct observations of atmospheric loss are needed to quantitatively test the predictions of various atmospheric evolutionary models. Several markers have been proposed and tested for tracing atmospheric escape \citep{2000ApJ...537..916S}. These include neutral hydrogen's Lyman $\alpha$ line in the UV \citep[e.g.,][]{ly_hd209458b, ly_gj3470b}, hydrogen's Balmer $\alpha$ (H\,$\alpha$) line \citep[e.g.,][]{Halpha_hd189_hd209, Halpha_hd189, aumic_teru}, UV metal lines \citep[e.g.,][]{uv_metal_hd209, uv_metal_GJ436b} and He\,I triplet lines at 10830\,\AA\ \citep[e.g.,][]{he_hatp11b_high, he_wasp107b_low}. He\,I observations are particularly interesting as they are not significantly absorbed by the interstellar medium (ISM) and hence are applicable to more distant worlds. Moreover, they can be observed from ground, even with a 3-meter class telescope \citep{oklopcic_hirata}. Due to its observability from the ground, it is the most commonly used evaporation marker, with over 50 exoplanet atmospheres probed using this marker. 

Helium exists in singlet or triplet states, based on the relative orientation of its electrons' spin. Any photon shortward of 504\,\AA\ can photoionize the helium atom. The subsequent recombination results in a transition to the ground state (1$\mathrm{^1}$S) for singlet helium. But for the triplet state, the lowest energy level (2$\mathrm{^3}$S) is decoupled from the singlet ground state (1$\mathrm{^1}$S) as the radiative transitions are suppressed. Hence the decay back to ground state takes over 2 hours, making 2$\mathrm{^3}$S a metastable state. The He\,I lines at 10830\,\AA\ arise from the 2$\mathrm{^3}$S--2$\mathrm{^3}$P metastable transition, producing three individual lines, two of which are blended and practically indistinguishable (10830.34\,\AA\ and 10830.25\,\AA) with a separate third line (10829.09\,\AA). They were primarily used for stellar \citep[e.g., chromospheric and stellar wind studies:][]{1992ApJ...387L..85D, 1994IAUS..154...35A} and galactic astrophysics \citep[e.g., quasars:][]{2011ApJ...728...94L}. In the exoplanet context, these lines were identified as promising markers for exoplanet atmospheres \citep{2000ApJ...537..916S}, especially the upper atmosphere \citep{2016MNRAS.458.3880T}. Although the search for helium in exoplanet atmospheres began in the early 2000s \citep[see][]{he_hd209_orig_2003}, it was not conclusively detected until recently in WASP-107b \citep{he_wasp107b_low}. Since then, over 50 planets have had their atmosphere probed at the 10830\,\AA\ He\,I feature. 

In this paper, we present a population-level approach to study patterns in the helium observed planets. The next section models the XUV flux received by the planet from its star. This is followed by a detailed section exploring our results. Finally, we conclude the paper with key takeaways in the conclusion section.

\section{Modeling} \label{modeling}

\subsection{Retrieving He I Papers from Archive}
We developed a semi-automatic pipeline to extract all the helium-related papers from \textit{arXiv} \citep{1994ComPh...8..390G}.We used the python wrapper {\tt arxiv}\footnote{\url{https://pypi.org/project/arxiv/}} package to access the papers from the database. This wrapper allows us to specify the period of search and keywords to narrow down the papers related to our study.  Specifically, we used keywords such as `helium', `He\,I', `10830\,\AA', and `observation' to shortlist the papers. The {\tt arxiv} package sifts through the title and abstract of the submitted articles for our keywords. This process helped us drastically reduce the number of papers that we manually sorted in the next step. 

Once we had the shortlisted papers, we manually sorted them to identify those with He\,I observations. Some of these planets have been observed multiple times \citep[e.g., WASP-107b:][]{he_wasp107b_low, he_wasp107b_high}, sometimes with opposing results \citep[e.g., WASP-52b:][]{he_wasp52b_wasp177b, allart_spirou}. In such cases, we used the results from the most recent study. We gathered all the results from the database until December 2023 and found that the He\,I triplet had been probed for 58 unique exoplanets (see Figure \ref{fig: mass-radius} for the radius--mass of all the planets).

We used {\tt exofile}\footnote{\url{https://github.com/AntoineDarveau/exofile}} to pull the stellar and planetary parameters from the NASA Exoplanet Archive. For properties that were not reported in the NASA Exoplanet Archive, we updated the parameters from the discovery papers. For planets that did not have their mass measured, we used the mass--radius relations from \cite{chen_kipping2017} to predict the mass (e.g., 7\,M$_{\oplus}$ for TOI-1430b). Similarly, we extrapolated the stellar rotation period from the v\,sin\,i for stars that did not have their rotation period measured. In all our figures, we under-plotted all the confirmed exoplanets with properties from the NASA Exoplanet Archive. The masses and radii of helium-observed planets, along with all the confirmed exoplanets, are plotted in Figure \ref{fig: mass-radius}. The pipeline is available in GitHub\footnote{\url{https://github.com/VigneshAstro/Helium-papers-from-arXiv}}.

\subsection{XUV flux estimation}

The X-ray and EUV luminosity of the star are not accurately estimated for all the planet-hosting stars \citep{2019sf2a.conf...59W}. Hence, to maintain uniformity in our analysis, we used scaling relations to estimate the X-ray and EUV flux at the planetary surface. Primarily, we obtained the X-ray (5--100\,\AA) and EUV (100--920\,\AA) luminosities, scaling them from the bolometric luminosity and age of the system using the relations in \cite{sanz_2011_x_euv_lum} (their equations 4, 5). We then calculated the planetary fluxes from the luminosities. Additionally, we computed the Lyman-$\alpha$ fluxes from the scaling relations in \cite{linsky2013_lya} by using a least square fit, log $f$(Ly$\alpha$) = A + B T$\mathrm{_{eff}}$, where T$\mathrm{_{eff}}$ is the effective temperature of the star. The parameters A and B depend on the rotation period of the host star \citep[see Section 6 and Table 5 of][]{linsky2013_lya}. This approach provides a good estimate for the Ly-$\alpha$ flux at 1\,AU from the host star, which we then re-scaled to the planet's orbital distance. We also estimated the cumulative X-ray + EUV (i.e., XUV) and Ly-$\alpha$ fluxes over the age of the planets. The cumulative XUV and Ly-$\alpha$ fluxes at the orbital distances of the planets are shown in Figure \ref{flux_plot}. We used the instantaneous XUV flux at the planet's orbit to estimate the energy-limited mass-loss. In the limit of perfect efficiency (i.e., $\epsilon = 1$):
\begin{equation}\label{eq:mdot}
    \dot{M} = \epsilon  \frac{\pi\, R_1^3 \, F_{XUV}}{G \, K \, M_p}
\end{equation} 
where $K$ (K\,$\le$\,1) accounts for the planet radius Roche lobe losses \citep{erkaev_2007_massloss} and G is the Newton's gravitational constant. To maintain uniformity, we assumed K = 1, as it is valid for most cases. In any case, this assumption yields only a lower limit of mass loss for all planets. Evaporation actually occurs at a point above the planetary radius $R_p$. But since most of the XUV radiation will be absorbed by material enclosed within the planet radius, we used the actual planetary radius for $R_1$.

\section{Results}

\subsection{Stellar XUV flux}

 We utilized scaling relations to estimate the X-ray and EUV luminosities following \cite{sanz_2011_x_euv_lum} and \cite{linsky2013_lya} since direct measurements for all target stars are exceedingly difficult. The mid-UV luminosity presents a similar challenge. However, no direct scaling relations exist to estimate mid-UV luminosities ($\lambda > 1000\,\mathrm{\AA}$). Some studies have employed metal lines like Mg\,II and \textit{GALEX} measurements to scale a portion of their mid-UV fluxes \citep{he_our_paper_gj9827}. Nevertheless, these metal line measurements and \textit{GALEX} data are unavailable for all exoplanet host stars. Another frequently used approach involves scaling the flux from the {\tt MUSCLES} database \citep{Muscles_lya} to reconstruct mid-UV fluxes for specific targets \citep[e.g., TOI-1807b:][]{he_toi1807b_toi2076b}. This method is suitable for low-mass stars, as the {\tt MUSCLES} database provides observed or reconstructed spectra for several M and K-type stars, but it is not applicable to hotter stars. For consistency, we employed scaled Ly-$\alpha$ ($\lambda = 1215.67\,\mathrm{\AA}$) flux as a proxy for the mid-UV flux. 
 
\begin{figure}[]
  \centering
  \includegraphics[width=\linewidth]{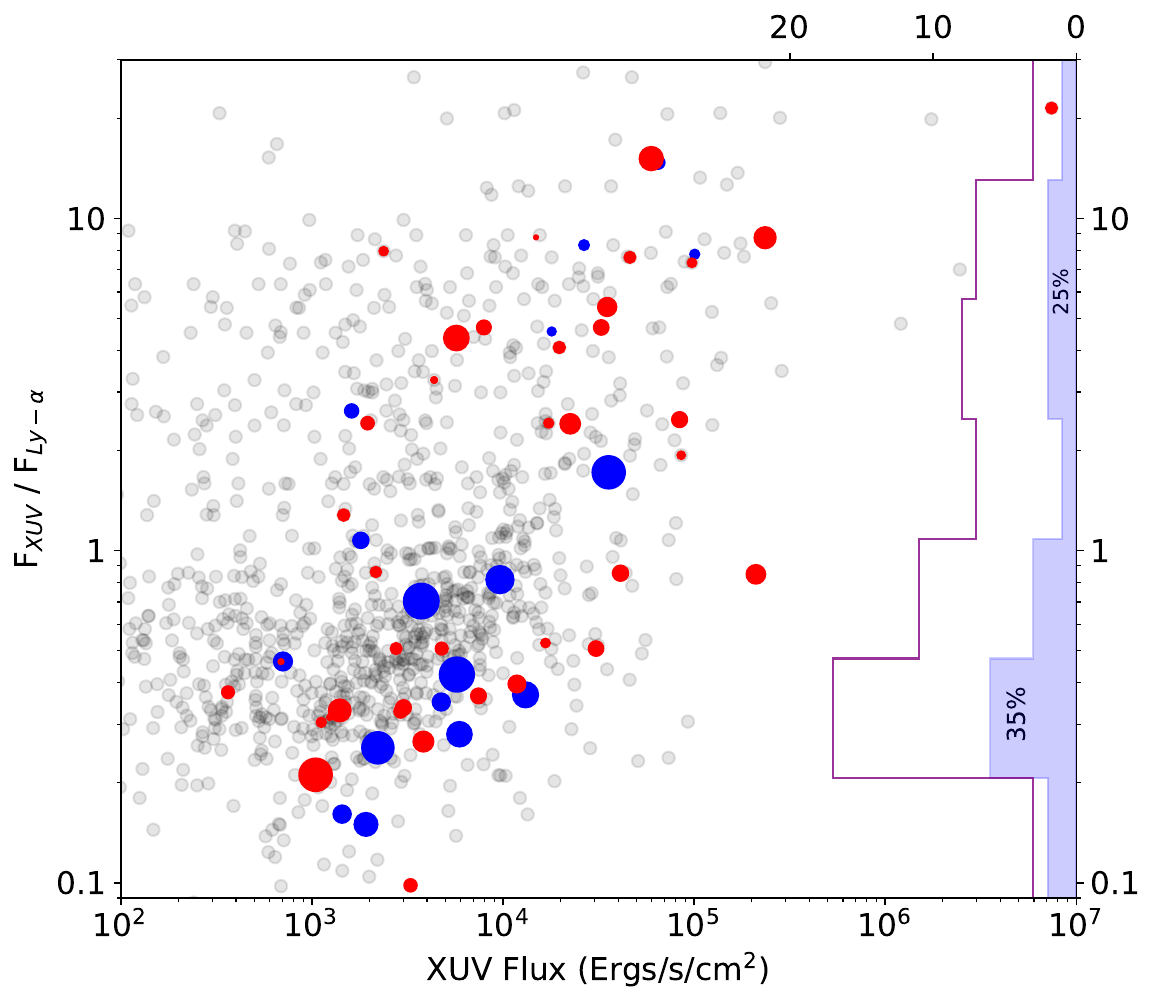}
  \caption{F$\mathrm{_{XUV}}$ / F$\mathrm{_{Ly-\alpha}}$ vs. F$\mathrm{_{XUV}}$ for all planets sourced from the NASA Exoplanet Archive. The blue and red TSM-scaled points denote the He\,I detected and undetected planets, respectively. The purple histogram shows the helium observed samples, and the accompanying histogram for detected planets (blue) illustrates that planets with detected He\,I tend to prefer lower F$\mathrm{_{XUV}}$ / F$\mathrm{_{Ly-\alpha}}$, contrary to predictions.}
  \label{xuv_vs_xuv_by_lyman}
\end{figure}

The high XUV flux arises from the youthfulness of the star and from stars that are highly active. Conversely, mid-UV flux is high for hotter stars and low for cooler stars. It is predicted that high XUV and low mid-UV fluxes are essential for populating helium in the metastable triplet state \citep{oklopcic_2019}. The ratio of XUV and mid-UV flux should be higher for detected planets. In Figure \ref{xuv_vs_xuv_by_lyman}, we plot the F$\mathrm{_{XUV}}$ / F$\mathrm{_{Ly-\alpha}}$ ratio with XUV flux. One can clearly see the preference for F$\mathrm{_{XUV}}$ / F$\mathrm{_{Ly-\alpha}}\,<\,$1 for most-detected planets over higher values, precisely the opposite of the expected trend. 

\subsection{Stellar effective temperature and metallicity}

\begin{figure}[]
  \centering
  \includegraphics[width=\linewidth]{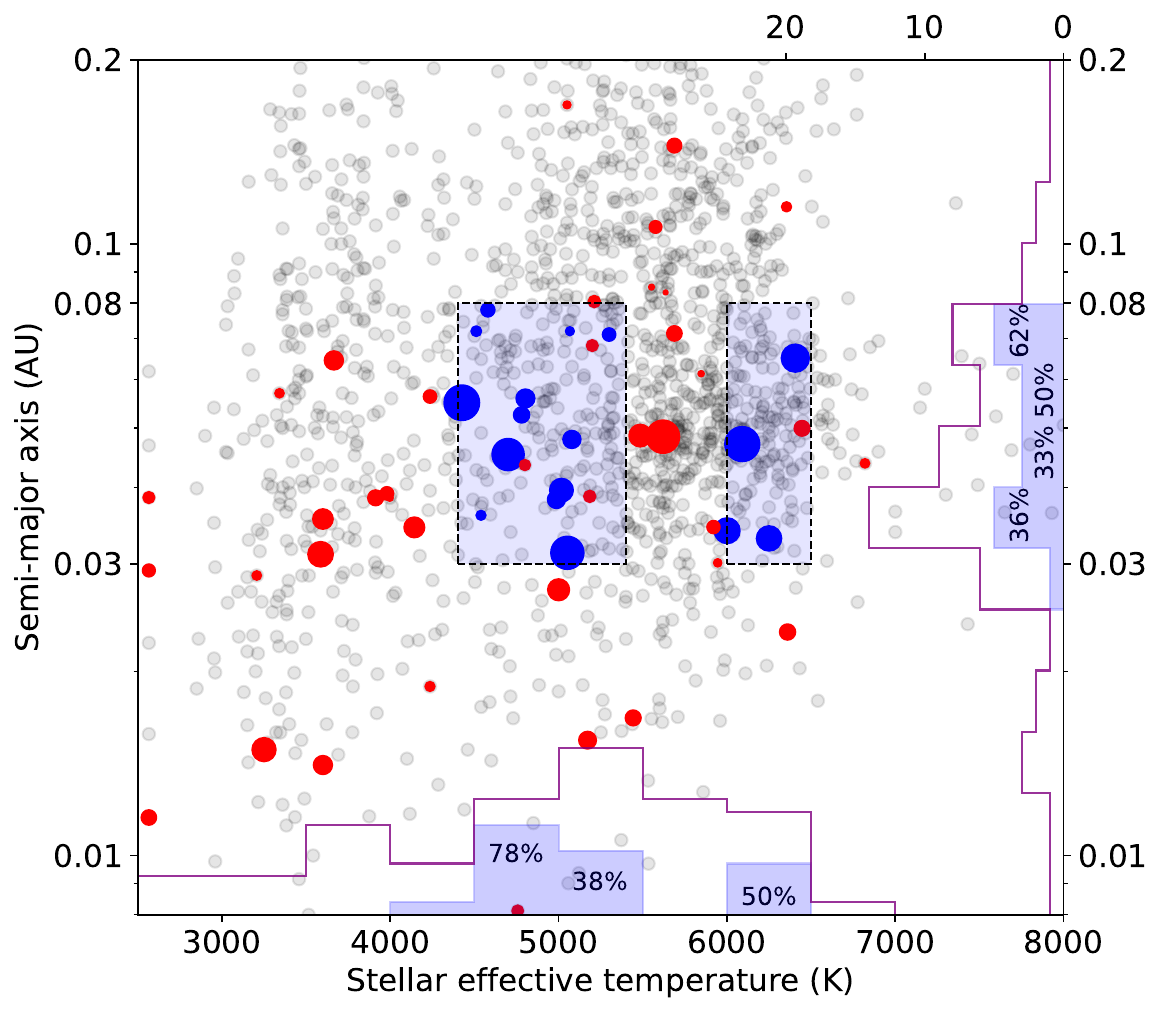}
  \caption{Semi-major axis vs. stellar effective temperature. The red and blue He\,I observed points are scaled similarly to those in Figure \ref{xuv_vs_xuv_by_lyman}. The `detection sweet spot', bounded by black boxes, encompasses the He\,I detected planets. The first (left) box contains K-star planets which have ideal spectra to excite He\,I, while the second box includes only giant planets around G-stars. Only hot Jupiters around G stars receive adequate flux and have sufficient helium to produce the He\,I feature.}
  \label{teff_semimajor}
\end{figure}

\begin{figure}[]
  \centering
  \includegraphics[width=\linewidth]{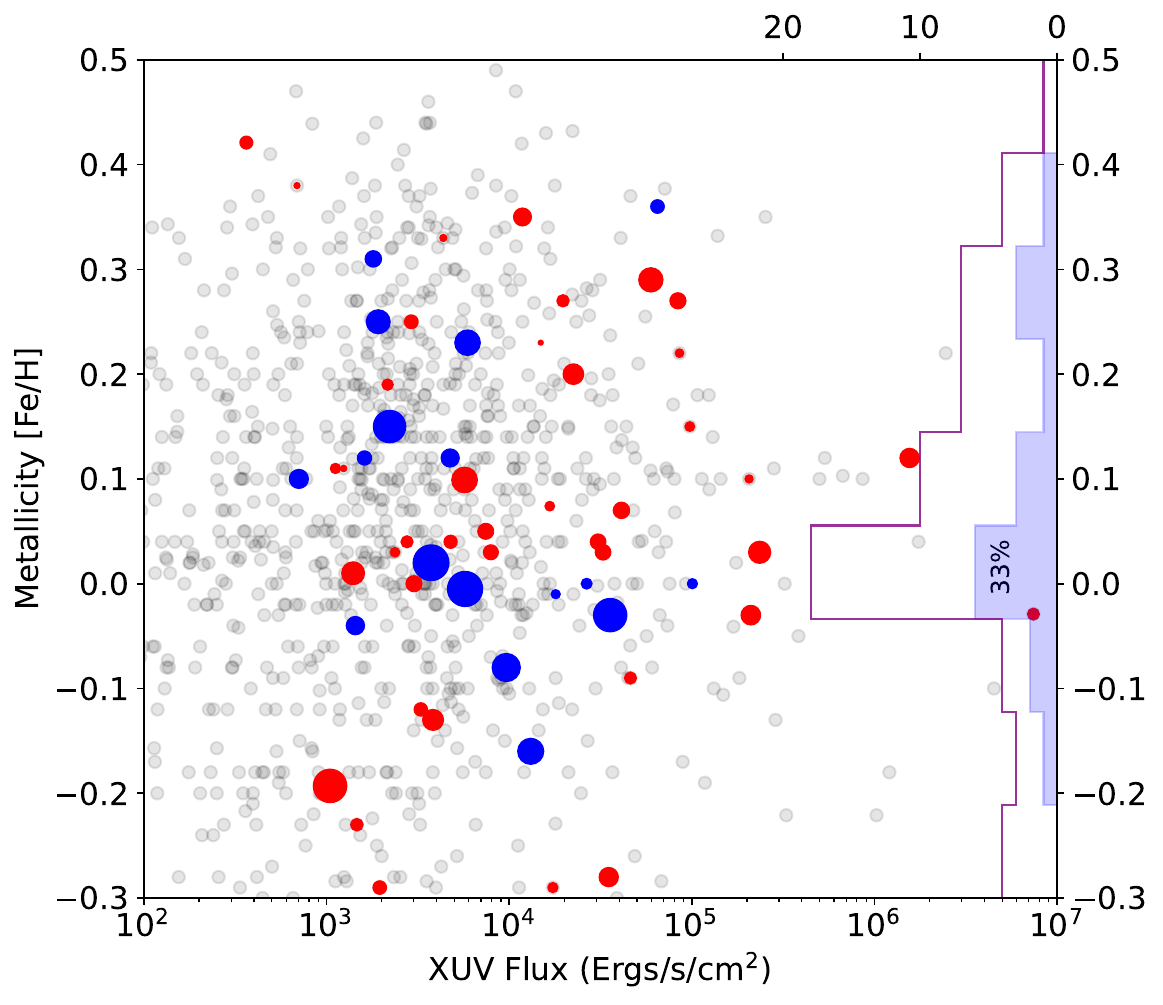}
  \caption{Stellar metallicity dependence on He\,I detection. The red and blue He\,I observed points are scaled similarly to those in Figure \ref{xuv_vs_xuv_by_lyman}. No preference in metallicity is observed.}
  \label{metallicity_plot}
\end{figure}

The star's spectral type and age play a crucial role in determining the XUV flux received by the planet. In Figure \ref{flux_plot}, we observe that both younger and older planets have been probed for helium. Although younger stars are more prone to higher stellar activity, thereby resulting in increased XUV luminosities \citep{oklopcic_2019}, they are not systematically preferred in detections. On the other hand, it can be seen that He\,I has been detected only around stars with T$\mathrm{_{eff}}$ between 4400--6500\,K (see Figure \ref{teff_semimajor}). This corresponds to stellar spectral type of K- and G-type stars, which were predicted to be the ideal spectral type to produce a sufficient population of helium in the metastable triplet state \citep{oklopcic_2019}. Peculiarly, we observe a gap in the stellar effective temperatures (5400\,K--6000\,K) where helium is not detected. 
The helium-detected planets show no preference with respect to metallicity (Figure \ref{metallicity_plot}). A slight increase in detections near solar metallicity could be due to observational biases or poor constraints on the star's metallicity.

\subsection{Planetary parameters}

Figure \ref{teff_semimajor} shows a distinct `\textit{detection sweet spot}' for helium detection, ranging from 0.03 AU to 0.08 AU and between 4400\,K and 6500\,K. According to \cite{oklopcic_2019}, planets within 0.05\,AU are anticipated to be conducive to He\,I triplet detections for K-type stars. For hotter stars, the orbit upper limit of the sweet spot should be further away due to the sheer magnitude of their XUV fluxes being too high at closer orbital distances. However, this trend is not clearly visible for hotter stars, possibly due to smaller sample sizes.

\begin{figure}[]
  \centering
  \includegraphics[width=\linewidth]{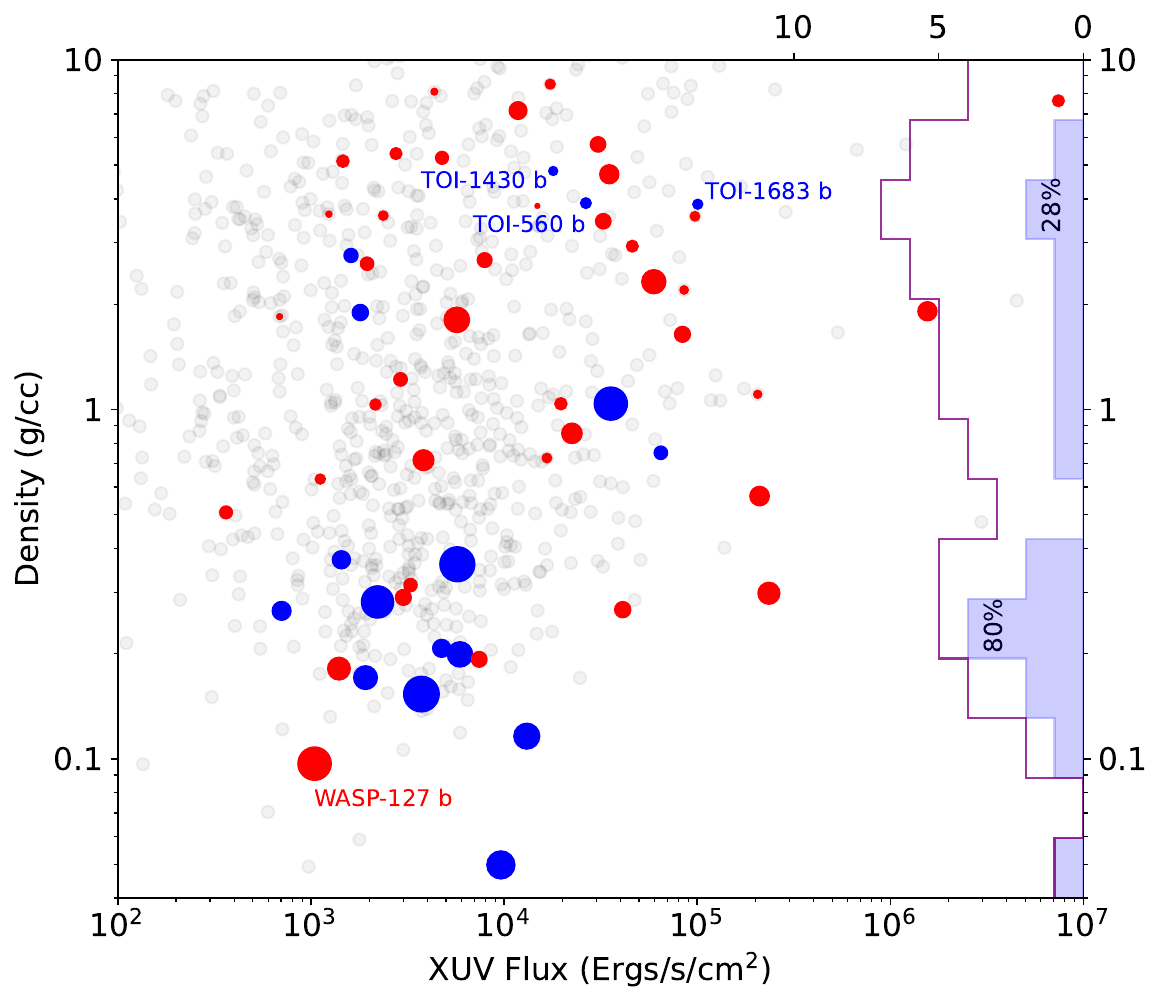}
  \caption{Planet density vs. F$\mathrm{_{XUV}}$ for all planets sourced from the NASA Exoplanet Archive. The red and blue He\,I observed points are scaled similarly to those in Figure \ref{xuv_vs_xuv_by_lyman}. The accompanying histograms illustrate that planets with detected He\,I (blue) tend to be lower density. Notorious planets are highlighted.}
  \label{density_plot}
\end{figure}

Close-in giant planets are the most amenable for He\,I detection as they have accreted a huge amount of primordial atmosphere while forming and receive high stellar influx and greater TSM. Even the strongest XUV irradiation cannot completely strip them of their atmosphere. Being less dense and hosting extended atmospheres makes them excellent targets for probing the exosphere. Evidently, they are the most helium-detected class of planets (see Figure \ref{density_plot}). On the other hand, atmospheres of smaller, high-density planets are difficult to detect as they might not have accumulated sufficient primordial atmosphere. Alternatively, even if they had, they might have already lost it through atmospheric escape. The non-detections on all planets could also be caused by insufficient stellar irradiation to drive the flow.

\section{Discussion}

\subsection{Stellar properties}

Ideally, we would use each exoplanet host star's own XUV, Ly-$\alpha$, and mid-UV fluxes for this analysis. However, obtaining such measurements for all stars would require thousands of hours on X-ray and UV telescopes, like India's \textit{AstroSat} \citep{astrosat_india}, NASA's \textit{XMM-Newton} \citep{xmm_newton} and \textit{Hubble Space Telescope} \citep{hubble_paper1}. Additionally, the measurements become even more challenging as the interstellar medium (ISM) absorption is high for distant host stars. Consequently, scaling relations become crucial for estimating XUV fluxes. Data products from the {\tt MUSCLES} survey are primarily targeted for planets around K- and M-type stars. However, for a comprehensive survey like ours, it may not be ideal, as scaling a K-star spectrum for an F- or G-type star could lead to erroneous results. Therefore, we utilized the scaling relations from \cite{sanz_2011_x_euv_lum} and \cite{linsky2013_lya}, as they can be extrapolated to a larger spectral range. Our He\,I observed samples span early A-type to late M-type star (see Figure \ref{teff_semimajor}). These models may not be accurate for stars hotter than F5V and colder than M5V. Moreover, the XUV values are model-dependent and are associated with stellar ages, which are not usually accurately constrained. Nonetheless, this estimation is far more accurate than other scaling methods in the literature. 

\begin{figure}[]
  \centering
  \includegraphics[width=\linewidth]{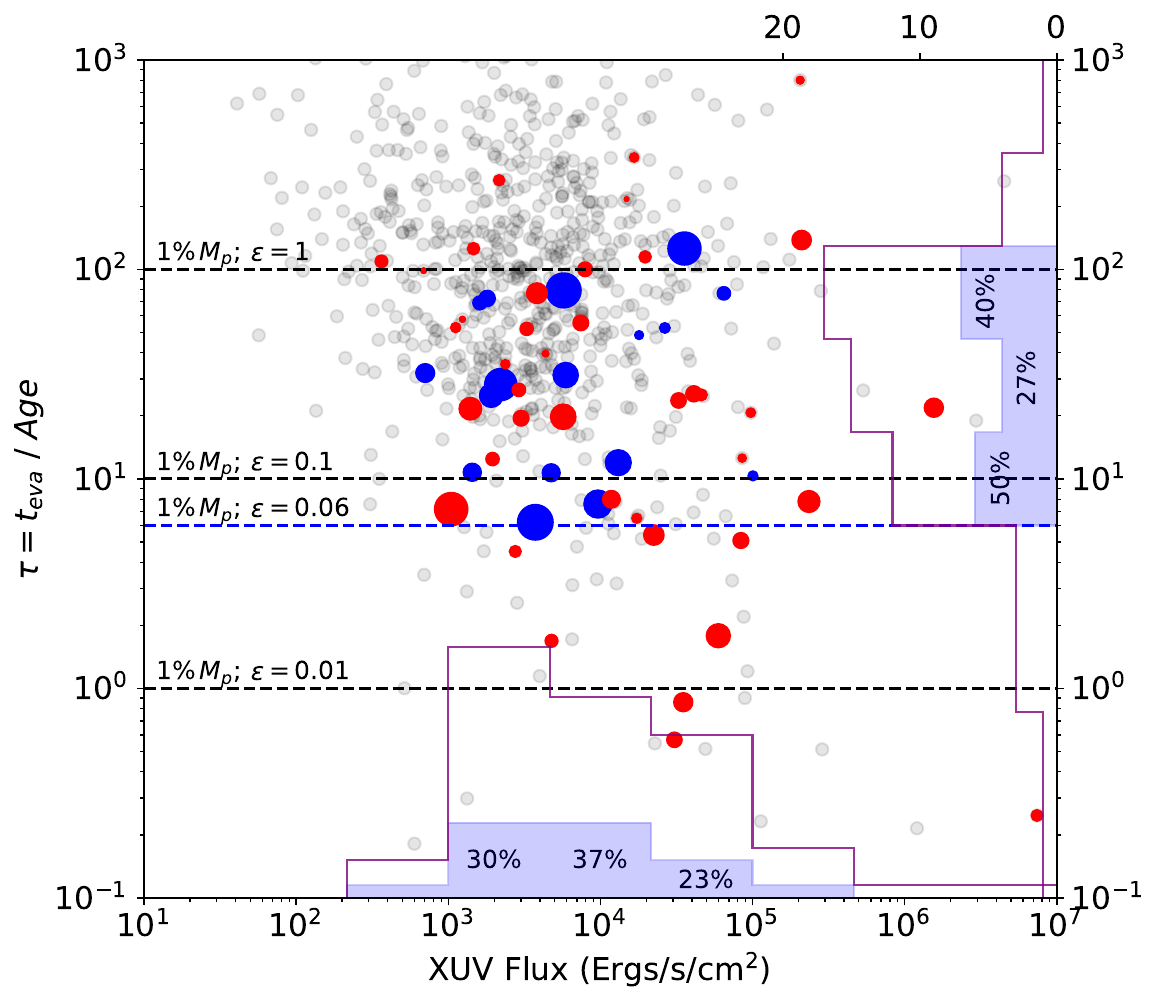}
  \caption{Expected atmospheric mass-loss vs. F$\mathrm{_{XUV}}$ for all planets sourced from the NASA Exoplanet Archive. The red and blue He\,I observed points are scaled similarly to those in Figure \ref{xuv_vs_xuv_by_lyman}. The horizontal black lines indicate the evaporation time (t$_{eva}$) equaling stellar age for different efficiencies ($\epsilon$), while the blue line indicates the lowest efficiency for the helium-detected planets. The accompanying histogram illustrates that planets with detected He\,I (blue) tend to prefer F$\mathrm{_{XUV}}$ flux in the range of $10^3$ to $10^5$ ergs/s/cm$^2$.}
  \label{fig: tau_xuv_plot}
\end{figure}

The preferred XUV flux falls within the range of $10^3$ to $10^5$ ergs/s/cm$^2$ (see Figure \ref{fig: tau_xuv_plot}). \cite{allart_spirou} arrived at a similar range (1400--17800 ergs/s/cm$^2$) with their SPIRou sample.  Although high XUV flux is expected to be preferred for populating the triplet state, the detections strongly lean towards the low XUV flux values (see Figure \ref{xuv_vs_xuv_by_lyman} and \ref{fig: tau_xuv_plot}). XUV flux must have been higher when these planets were younger. This, combined with the high density of planets (see Figure \ref{density_plot}), could explain the non-detections for planets with high XUV. Additionally, we computed energy-limited mass-loss rates for these planets to estimate the time to evaporate the complete atmosphere (t$_{eva}$): 

\begin{equation} \label{eqn2}
    t_{eva} = \frac{M_{p}}{\dot{M}} = \frac{M_p}{\epsilon \frac{\pi \, R_1^3 \, F_{XUV}}{G \, K \, M_p}}
\end{equation}
where, M$_{p}$ is the planetary mass and $\dot{M}$ is the energy-limited mass-loss rate seen in equation \ref{eq:mdot}. We define the ratio $\tau$ = t$_{eva}$ / age as a parameter to understand the effect of heating efficiency on evaporation. Here, we are not considering complete planet evaporation but only the atmosphere; hence, we assume a nominal 1\% of the planet's mass as atmospheric mass.  Although this is a fair assumption \citep{owen_wu_2016, 2017ApJ...847...29O}, planets, especially giant planets, might have accreted more gas while forming. We also used the instantaneous XUV flux in these calculations. For any planet located above the black dashed lines in Figure \ref{fig: tau_xuv_plot}, they have t$_{eva}\,>$ age, indicating that the planet is not old enough to have its atmosphere completely evaporated. On the other hand, planets located below the dashed lines have t$_{eva}\,<$ age, suggesting that their atmospheres might have been completely evaporated by the time we observed them. The position of this transition line strongly depends on how effectively the planets utilize the XUV flux for the mass-loss process. We report a lower-limit efficiency of 6\% ($\epsilon$=0.06) for all the helium-detected planets.

\subsection{Detection sweet spot--cosmic shoreline}

\cite{oklopcic_2019} predicted that K-type stars are preferred for populating helium in the metastable triplet state. Observed patterns also support this prediction (Figure \ref{teff_semimajor}), with all detections happening around stars with effective temperatures of 4400--6500\,K. However, the distribution is not uniform, as we see a gap in the stellar effective temperatures between 5400\,K and 6000\,K for helium-detected planets. The first group of detected planets (left box with detected planets in Figure \ref{teff_semimajor}) all orbit around K-type stars, supporting the predictions. However, the second group of planets all orbit G-type stars, and all these planets are giants. The strong XUV flux combined with sufficient helium in the atmospheres of these hot Jupiters are probable causes for these detections. Furthermore, the non-detections around low-mass stars can be attributed to their smaller protoplanetary disks, implying smaller amounts of gas available for accretion \citep{lee_chiang_ormel2014, lee_chiang2016}.

All the helium-detected planets lie between the semi-major axis distance of 0.03\,AU and 0.08\,AU. The smaller planets situated much closer than 0.03\,AU may have lost their primordial atmospheres within a few thousand to a million years due to the extreme XUV flux received from their host stars \citep{2017ApJ...847...29O}. Unless we observe these planets in their ultra-young ages, detections are not expected. Similarly, smaller planets in more distant orbits may not be detected due to the current low XUV flux they receive, in contrast to the potentially high XUV flux they experienced in their youth. 

In the case of giant planets, there are no Jupiter-sized planets observed below 0.03\,AU with probed He\,I except for WASP-52b and WASP-80b. For WASP-52b, \cite{he_wasp52b_wasp177b} reported a strong absorption signal in their Keck/NIRSPEC data. However, \cite{allart_spirou} argued that the positive detection observed by \cite{he_wasp52b_wasp177b} could be a pseudo-stellar signal resulting from inhomogeneous stellar surface occultations, as their CFHT/SPIRou data did not show any detection. Furthermore, WASP-52 is an old star with reduced XUV luminosity \citep{wasp52b_age}, potentially explaining its non-detection. Conversely, although WASP-80b is young \citep{wasp80b_age}, its high X-ray flux and very low EUV flux make it a poor candidate for producing a detectable He\,I absorption signal \citep{wasp80b_hemodel}. The closest giant planet for which we see a positive detection is HD 189733b \citep{he_hd189733b_1}. Similar to small planets, the giant planet further from the host star does not receive enough XUV flux to excite the helium atoms to produce the He\,I feature. 

\begin{figure}[]
  \centering
  \includegraphics[width=\linewidth]{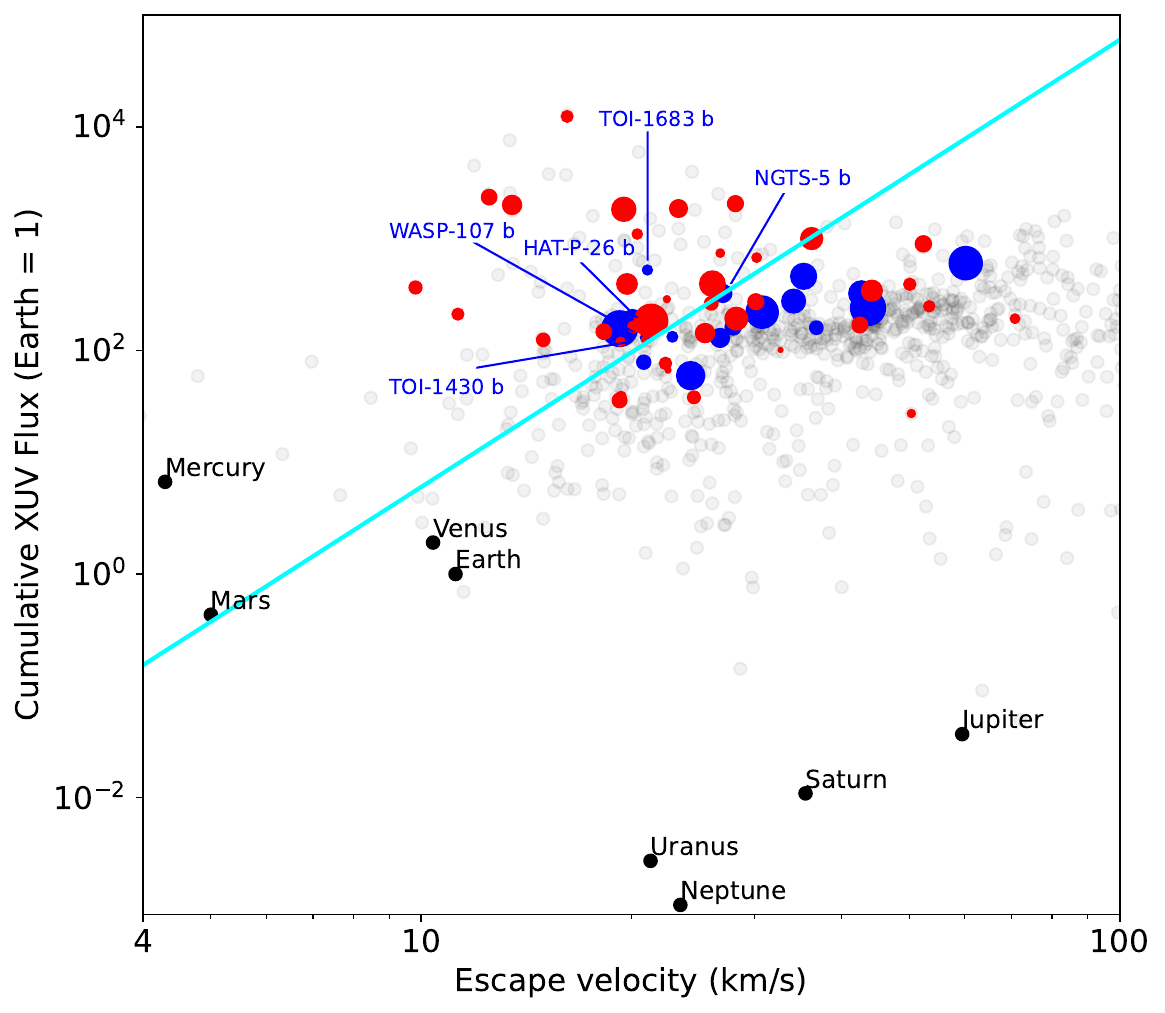}
  \caption{The cumulative XUV flux (normalized with Earth) with the escape velocity of the planets. The \textit{cosmic shoreline} ($ I\mathrm{_{XUV}} \propto v\mathrm{_{esc}^4}$), shown in cyan, is reconstructed from \cite{cosmic_shoreline}. The planets below this line are expected to retain their atmosphere, while those above are expected to deplete theirs. Notorious planets above the cosmic shoreline are labeled.}
  \label{fig: cosmic_shoreline}
\end{figure}

We can also examine the same planets in the accumulated XUV flux vs. escape velocity (see Figure \ref{fig: cosmic_shoreline}). We plot the `cosmic shoreline' \citep{cosmic_shoreline}, which uses a simple $ I\mathrm{_{XUV}} \propto v\mathrm{_{esc}^4}$ power law to differentiate planets that can retain an atmosphere (right of the line) from the ones that do not. Although a significant number of helium-detected planets lie below the shoreline, a few lie on above too. These planets might completely evaporate their atmosphere through XUV-driven processes if they are small. Notably young mini-Neptunes like TOI-1430b \citep{he_zhang_2023_4minineptuens, he_toi1430b_new} and  TOI-1683 b \citep{he_shreyas_neptunes} which are evaporating would eventually completely strip their primordial atmospheres and become super-Earths. But giant planets like WASP-107b \citep{he_wasp107b_low, he_wasp107b_high} and NGTS-5b \citep{he_shreyas_neptunes} and Neptune-sized planet HAT-P-26b \citep{he_shreyas_neptunes} might retain their atmosphere, despite their current mass-loss. We note that these planets with atmospheres on the wrong side of the shoreline could mean that our historical XUV estimates are high compared to the actual values. Alternatively, the `cosmic shoreline' itself could be incorrect, and there could be a different mechanism at play, such as core-powered mass loss \citep{2018MNRAS.476..759G}, driving the outflow.

\subsection{Small planets vs. Large planets}

All planets are expected to be born with a primordial hydrogen-helium atmosphere \citep{2013ApJ...775..105O}. Younger planets are irradiated with significant higher XUV luminosity from the star compared to their older counterparts. If the planet is small ($R_p< \,2.6\,R_{\oplus}$), then even if it had accreted $\sim$1\% by mass H/He primordial atmosphere, it would be completely stripped of its atmosphere \citep{owen_wu_2016, 2017ApJ...847...29O}. Hence, non-detections from older systems with smaller planets like the TRAPPIST-1 system make complete sense \citep{he_trappist1_our_paper}. Therefore, unless we observe them extremely young and during the process of evaporation, we may not detect any escaping helium. This is precisely why we do not detect helium in any older smaller planets (right plot of Figure \ref{flux_plot}). In the density plot (Figure \ref{density_plot}), we see that smaller planets TOI-560b, TOI-1430b, and TOI-1683b are experiencing helium escape. These are young planets, and we are precisely capturing the evaporation as it happens. In the case of large planets we expect detections across the age and not just in young hot giants. However the non-detections in these giants can be explained by the low XUV flux the planet receives. For example WASP-127b comparatively receives little XUV flux due to its old age of the system \citep[11.40\,$\pm$\,1.8\,Gyr;][]{he_wasp127b}. 

\section{Conclusion}

In this study, we explored the population-level patterns of helium detections and non-detections in exoplanet atmospheres and investigated the underlying factors influencing these detections. Here are the key takeaways from our investigation:
\begin{itemize}
    \item \textbf{Stellar Properties:} As predicted by \cite{oklopcic_2019}, the effective temperature of the host star plays a crucial role in helium detection, with most detections occurring around K- and G-type stars with effective temperatures between 4400--5400\,K and 6000--6500\,K. We also observe a gap in the stellar effective temperature range (5400--6000\,K) where no detections occur. Additionally, helium detected planets does not show a significant correlation with the stellar metallicity.
    \item \textbf{XUV Flux:} We report a noteworthy trend where helium-detected planets prefer a lower ratio of XUV flux to mid-UV flux (F$\mathrm{_{XUV}}$ / F$\mathrm{_{Ly-\alpha}}$ $<$ 1) compared to non detections. Among the helium probed planets with low F$\mathrm{_{XUV}}$ / F$\mathrm{_{Ly-\alpha}}$, 35\% have detected helium in their atmosphere, while only 25\% of planets detect helium for higher ratios. Further investigations are needed to understand this observed trend.
    \item \textbf{Orbital Sweet Spot:} Our analysis revealed a distinct sweet spot for helium detection, ranging from 0.03\,AU to 0.08\,AU, where all the detections occur. Our analysis of the `cosmic shoreline' \citep{cosmic_shoreline} shows planets exhibiting atmospheric escape do exist above the shoreline, suggesting an improved predictions on the shoreline is needed.
    \item \textbf{Small vs. Large Planets:} Smaller planets closer to their host stars are more prone to atmospheric escape due to intense XUV irradiation, resulting in non-detections in older systems. On the other hand, large planets may retain their atmospheres even in older systems if the XUV flux is low.
\end{itemize}

In conclusion, our study provides valuable insights into the complex interplay between stellar properties, orbital dynamics, and atmospheric escape mechanisms in exoplanetary systems. Further understanding of the dependence of atmospheric escape on the XUV flux received by the planet is needed. The absence of helium-detected planets between 5400--6000\,K highlights the importance of considering stellar and orbital parameters beyond just the XUV flux.

\begin{acknowledgements}
VK thanks Antoine Darveau-Bernier for helping him understand the {\tt exofile} package. VK also thanks Romain Allart and Teruyuki Hirano for their general review on this topic. Authors thanks Eric Gaidos and Katherine Bennett for their fruitful conversations during the Exoclimes VI conference.
\facility {Exoplanet Archive, arxiv, exofile}

\end{acknowledgements}

\bibliography{Helium_vignesh_ApJ}{}
\bibliographystyle{aasjournal}

\end{document}